\documentclass[usegraphicx, longauth]{aa}

\usepackage{natbib}
\bibpunct{(}{)}{;}{a}{}{,} 
\usepackage{subfigure}
\usepackage{longtable}
\usepackage{graphicx}
\usepackage{graphics}
\usepackage{keyval}
\usepackage{trig}
\usepackage[american]{babel}
\usepackage{url}
\usepackage{color}
\usepackage{amsmath}
\usepackage{bm}
\usepackage{longtable}

\def\beq{\begin{equation}}
\def\eeq{\end{equation}}
\def\bey{\begin{eqnarray}}
\def\eey{\end{eqnarray}}

\def\msun{M_\odot}

\def\lsim{\mathrel{\raise.3ex\hbox{$<$\kern-.75em\lower1ex\hbox{$\sim$}}}}
\def\gsim{\mathrel{\raise.3ex\hbox{$  $\kern-.75em\lower1ex\hbox{$\sim$}}}}
\def\Msun{M_\odot}

\def\tz{t_\mathrm{0}}
\def\te{t_\mathrm{E}}

\def\uz{u_\mathrm{0}}
\def\Mjup{M_\mathrm{J}}

\def\thetae{\theta_\mathrm{E}}
\def\fs{F_\mathrm{S}}
\def\fb{F_\mathrm{B}}
\def\ds{D_\mathrm{S}}
\def\dl{D_\mathrm{L}}

\def\a0{A_{\mathrm{0}}}

\def\pie{\pi_{E}}
\def\piee{\pi_{E, E}}
\def\pien{\pi_{E, N}}

\def\uas{\mu \mathrm{as}}

\def\apjl{ApJ}

\def\aap{A\&A}      
\def\apjs{ApJS}       

\newcommand\Eq[1]{Eq.~(\ref{#1})}
\newcommand\Fig[1]{Fig.~\ref{#1}}
\newcommand\Tab[1]{Table~\ref{#1}}
\newcommand\Sec[1]{Sec.~\ref{#1}}

\begin{document}
   \title{A Giant Planet beyond the Snow Line in Microlensing Event OGLE-2011-BLG-0251}

   \author{N. Kains
          \inst{\ref{eso}}\fnmsep\thanks{corresponding author; nkains@eso.org}
          \and
          R. A. Street\inst{\ref{lcogt}}
          \and
          J.-Y. Choi \inst{\ref{chungbuk}} \and
          C. Han\inst{\ref{chungbuk}}\fnmsep\thanks{corresponding author; cheongho@astroph.chungbuk.ac.kr}
          \and
          A. Udalski\inst{\ref{warsaw}}\and
          L.~A. Almeida\inst{\ref{saojose}} \and 
	F.~Jablonski\inst{\ref{saojose}} \and
	P.J.~Tristram\inst{\ref{uvic}}\and
 	U.G. J{\o}rgensen \inst{\ref{nbi},\ref{copenhagen}} \and
         \\
          and \\
          M.\,K. Szyma{\'n}ski\inst{\ref{warsaw}} \and 
	M. Kubiak\inst{\ref{warsaw}} \and G.
	Pietrzy{\'n}ski\inst{\ref{warsaw}, \ref{concepcion}} \and 
	I. Soszy{\'n}ski\inst{\ref{warsaw}} \and 
	R. Poleski\inst{\ref{warsaw}} \and
	S. Koz{\l}owski\inst{\ref{warsaw}} \and 
	P. Pietrukowicz\inst{\ref{warsaw}} \and 
	K. Ulaczyk\inst{\ref{warsaw}} \and
	{\L}. Wyrzykowski\inst{\ref{ioa}, \ref{warsaw}} \and 
	J. Skowron\inst{\ref{osu}, \ref{warsaw}}\\
	(The OGLE collaboration)\\
	and\\
          K.A. Alsubai \inst{\ref{qnrf}} \and
	V. Bozza \inst{\ref{salerno}, \ref{fisicanucleare}} \and
	P. Browne \inst{\ref{standrews}} \and
	M.J. Burgdorf \inst{\ref{sofiade}, \ref{sofia_ames}} \and
	S. Calchi Novati \inst{\ref{salerno}, \ref{iiass}} \and
	P. Dodds \inst{\ref{standrews}} \and
	M. Dominik \inst{\ref{standrews}}\fnmsep\thanks{Royal Society University Research Fellow} \and
	S. Dreizler \inst{\ref{gottingen}} \and
	X.-S. Fang \inst{\ref{naochina}} \and
	F. Grundahl \inst{\ref{aarhus}} \and
	C.-H. Gu \inst{\ref{naochina}} \and
	S. Hardis \inst{\ref{nbi}} \and
	K. Harps{\o}e \inst{\ref{nbi}, \ref{copenhagen}} \and
	F.V. Hessman \inst{\ref{gottingen}} \and
	T.C. Hinse \inst{\ref{kasi},\ref{nbi}, \ref{armagh}} \and
	A. Hornstrup \inst{\ref{manchester}} \and
	M. Hundertmark \inst{\ref{standrews},\ref{gottingen}} \and
	J. Jessen-Hansen \inst{\ref{aarhus}} \and
	E. Kerins \inst{\ref{manchester}} \and
	C. Liebig \inst{\ref{standrews}} \and
	M. Lund \inst{\ref{aarhus}} \and
	M. Lundkvist \inst{\ref{aarhus}} \and
	L. Mancini \inst{\ref{mpia}} \and
	M. Mathiasen \inst{\ref{nbi}} \and
	M.T. Penny \inst{\ref{manchester}, \ref{osu}} \and
	S. Rahvar \inst{\ref{sharif}, \ref{perimeter}} \and
	D. Ricci \inst{\ref{liege}} \and
	K.C. Sahu \inst{\ref{stsci}} \and
	G. Scarpetta \inst{\ref{salerno},\ref{infn}} \and
	J. Skottfelt \inst{\ref{nbi}} \and
	C. Snodgrass \inst{\ref{mps}} \and
	J. Southworth \inst{\ref{keele}} \and
	J. Surdej \inst{\ref{liege}} \and
	J. Tregloan-Reed \inst{\ref{keele}} \and
	J. Wambsganss \inst{\ref{ari}} \and
	O. Wertz \inst{\ref{liege}} \\
          (The MiNDSTEp consortium)\\
          and\\
          D. Bajek \inst{\ref{standrews}} \and
          D.M.~Bramich\inst{\ref{eso}} \and
          K.~Horne\inst{\ref{standrews}} \and 
          S. Ipatov \inst{\ref{alsubai}} \and
          I.A.~Steele\inst{\ref{ljmu}} \and
          Y.~Tsapras\inst{\ref{lcogt}, \ref{queenmary}}\\
          (The RoboNet collaboration)\\
	and\\
	F.~Abe\inst{\ref{nagoya}}\and
	D.P.~Bennett\inst{\ref{notredame}}\and 
	I.A.~Bond\inst{\ref{massey}}\and 
	C.S.~Botzler\inst{\ref{auckland}}\and 
	P. Chote\inst{\ref{uvic}}\and 
	M. Freeman\inst{\ref{auckland}}\and 
	A.~Fukui\inst{\ref{okayama}}\and 
	K.~Furusawa\inst{\ref{nagoya}}\and
	Y.~Itow\inst{\ref{nagoya}}\and 
	C.H.~Ling\inst{\ref{massey}}\and 
	K.~Masuda\inst{\ref{nagoya}}\and 
	Y.~Matsubara\inst{\ref{nagoya}}\and 
	N.~Miyake\inst{\ref{nagoya}}\and
	Y.~Muraki\inst{\ref{konan}}\and 
	K.~Ohnishi\inst{\ref{nagano}}\and 
	N.~Rattenbury\inst{\ref{auckland}}\and 
	T.~Saito\inst{\ref{tokyomet}}\and
	D.J.~Sullivan\inst{\ref{uvic}}\and
	T. Sumi\inst{\ref{osaka}}\and 
	D.~Suzuki\inst{\ref{osaka}}\and 
	K. Suzuki\inst{\ref{nagoya}}\and 
	W.L.~Sweatman\inst{\ref{massey}}\and 
	S. Takino\inst{\ref{nagoya}}\and 
	K.~Wada\inst{\ref{osaka}}\and 
	P.C.M.~Yock\inst{\ref{auckland}}\\
	(The MOA collaboration)\\	
	and\\
	W. Allen\inst{\ref{vintagelane}}\and
	V. Batista \inst{\ref{osu}} \and
	S.-J. Chung \inst{\ref{kasi}} \and
	G. Christie\inst{\ref{aucklandobs}}\and
	D.L.~DePoy\inst{\ref{texasam}} \and
	J. Drummond\inst{\ref{possum}}\and
	B.S.~Gaudi\inst{\ref{osu}} \and
	A. Gould \inst{\ref{osu}} \and
	C. Henderson \inst{\ref{osu}} \and
	Y.-K. Jung\inst{\ref{chungbuk}}\and
	J.-R. Koo\inst{\ref{kasi}} \and
	C.-U.~Lee\inst{\ref{kasi}} \and 
	J. McCormick \inst{\ref{farmcove}}\and
	D. McGregor \inst{\ref{osu}}\and
	J.A.~Mu{\~n}oz\inst{\ref{valencia}} \and
	T. Natusch \inst{\ref{aucklandobs}, \ref{aucklandaut}}\and
	H. Ngan  \inst{\ref{aucklandobs}}\and
	H. Park\inst{\ref{chungbuk}}\and
	R.W.~Pogge\inst{\ref{osu}} \and 
	I.-G.~Shin\inst{\ref{chungbuk}} \and 
	J.~Yee\inst{\ref{osu}} \\
         (The $\mu$FUN collaboration)\\
         and\\
	M.D.~Albrow\inst{\ref{canterbury}} \and
	E.~Bachelet\inst{\ref{latt}, \ref{cnrstoulouse}} \and
	J.-P.~Beaulieu\inst{\ref{iap}} \and 
	S.~Brillant\inst{\ref{esochile}} \and
	J.A.R.~Caldwell\inst{\ref{mcdonald}} \and 
	A.~Cassan\inst{\ref{iap}} \and
	A.~Cole\inst{\ref{utas}} \and 
	E.~Corrales\inst{\ref{iap}} \and
	Ch.~Coutures\inst{\ref{iap}} \and 
	S.~Dieters\inst{\ref{latt}} \and
	D.~Dominis Prester\inst{\ref{rijeka}} \and 
	J.~Donatowicz\inst{\ref{tuv}} \and 
	P.~Fouqu{\'e}\inst{\ref{latt}, \ref{cnrstoulouse}} \and
	J.~Greenhill\inst{\ref{utas}} \and
	S.R.~Kane\inst{\ref{nexsci}} \and 
	D.~Kubas\inst{\ref{esochile}, \ref{iap}} \and 
	J.-B.~Marquette\inst{\ref{iap}} \and
	R.~Martin\inst{\ref{perthobs}} \and 
	P.~Meintjes\inst{\ref{freestate}} \and 
	J.~Menzies\inst{\ref{saao}} \and
	K.R.~Pollard\inst{\ref{canterbury}} \and 
	A.~Williams\inst{\ref{ari}} \and
	D.~Wouters\inst{\ref{iap}} \and 
	M.~Zub\inst{\ref{ari}}\\
	(The PLANET collaboration)\\
          }

\institute{
European Southern Observatory, Karl-Schwarzschild Stra\ss e 2, 85748 Garching bei M\"{u}nchen, Germany \label{eso}\\
\and Las Cumbres Observatory Global Telescope Network, 6740 Cortona Drive, Suite 102, Goleta, CA 93117, USA  \label{lcogt}\\	
\and Department of Physics, Institute for Astrophysics, Chungbuk National University, Cheongju 371-763, Korea \label{chungbuk}\\
\and Warsaw University Observatory, Al. Ujazdowskie 4, 00-478 Warszawa, Poland \label{warsaw}\\	
\and Divis\~{a}o de Astrofisica, Instituto Nacional de Pesquisas Espeaciais, Avenida dos Astronauntas, 1758 Sao Jos{\'e} dos Campos, 12227-010 SP, Brazil \label{saojose} \\  
\and School of Chemical and Physical Sciences, Victoria University, Wellington, New Zealand \label{uvic}\\
\and Niels Bohr Institute, University of Copenhagen, Juliane Maries vej 30, 2100 Copenhagen, Denmark \label{nbi}\\	
\and Centre for Star and Planet Formation, Geological Museum, {\O}ster Voldgade 5, 1350 Copenhagen, Denmark \label{copenhagen}\\	
\and Qatar Foundation, P.O. Box 5825, Doha, Qatar \label{qnrf}\\	
\and Dipartimento di Fisica ``E.R Caianiello", Universitˆ di Salerno, Via Ponte Don Melillo, 84084 Fisciano, Italy \label{salerno}\\	
\and Istituto Nazionale di Fisica Nucleare, Sezione di Napoli, Italy \label{fisicanucleare}\\
\and SUPA School of Physics \& Astronomy, University of St Andrews, North Haugh, St Andrews, KY16 9SS, United Kingdom \label{standrews}\\	
\and Deutsches SOFIA Institut, Universit\"{a}t Stuttgart, Pfaffenwaldring 31, 70569 Stuttgart, Germany \label{sofiade}\\	
\and SOFIA Science Center, NASA Ames Research Center, Mail Stop N211-3, Moffett Field CA 94035, United States of America \label{sofia_ames}\\	
\and Istituto Internazionale per gli Alti Studi Scientifici (IIASS), Vietri Sul Mare (SA), Italy \label{iiass}\\		
\and Institut f\"{u}r Astrophysik, Georg-August-Universit\"{a}t, Friedrich-Hund-Platz 1, 37077 G\"{o}ttingen, Germany \label{gottingen}\\	
\and Korea Astronomy and Space Science Institute, Daejeon 305-348, Korea \label{kasi}\\
\and National Astronomical Observatories/Yunnan Observatory, Joint laboratory for Optical Astronomy, Chinese Academy of Sciences, Kunming 650011, People's Republic of China \label{naochina}\\	
\and Department of Physics and Astronomy, Aarhus University, Ny Munkegade 120, 8000 {\AA}rhus C, Denmark \label{aarhus}\\	
\and Armagh Observatory, College Hill, Armagh, BT61 9DG, Northern Ireland, United Kingdom \label{armagh}\\	
\and Danmarks Tekniske Universitet, Institut for Rumforskning og -teknologi, Juliane Maries Vej 30, 2100 K{\o}benhavn, Denmark \label{dtu}\\	
\and Jodrell Bank Centre for Astrophysics, University of Manchester, Oxford Road,Manchester, M13 9PL, UK \label{manchester}\\	
\and Max Planck Institute for Astronomy, K\"{o}nigstuhl 17, 69117 Heidelberg, Germany \label{mpia}\\	
\and Department of Astronomy, Ohio State University, 140 West 18th Avenue, Columbus, OH 43210, United States of America \label{osu}\\	
\and Department of Physics, Sharif University of Technology, P.~O.\ Box 11155--9161, Tehran, Iran \label{sharif}\\	
\and Perimeter Institute for Theoretical Physics, 31 Caroline St. N., Waterloo ON, N2L 2Y5, Canada \label{perimeter}\\	
\and Institut d'Astrophysique et de G\'{e}ophysique, All\'{e}e du 6 Ao\^{u}t 17, Sart Tilman, B\^{a}t.\ B5c, 4000 Li\`{e}ge, Belgium \label{liege}\\	
\and Space Telescope Science Institute, 3700 San Martin Drive, Baltimore, MD 21218, United States of America \label{stsci}\\
\and INFN, Gruppo Collegato di Salerno, Sezione di Napoli, Italy \label{infn}\\	
\and European Southern Observatory (ESO), Alonso de Cordova 3107, Casilla 19001, Santiago 19, Chile \label{esochile}\\	
\and Max Planck Institute for Solar System Research, Max-Planck-Str. 2, 37191 Katlenburg-Lindau, Germany \label{mps}\\	
\and Astrophysics Group, Keele University, Staffordshire, ST5 5BG, United Kingdom \label{keele}\\	
\and Astronomisches Rechen-Institut, Zentrum f\"{u}r Astronomie der Universit\"{a}t Heidelberg (ZAH),  M\"{o}nchhofstr.\ 12-14, 69120 Heidelberg, Germany \label{ari}\\
\and Institute of Astronomy, University of Cambridge, Madingley Road, Cambridge CB3 0HA, United Kingdom \label{ioa}\\
\and AlsubaiÕs Establishment for Scientific Studies, Doha, Qatar \label{alsubai}\\
\and Astrophysics Research Institute, Liverpool John Moores University, Twelve Quays House, Egerton Wharf, Birkenhead, Wirral., CH41 1LD, UK \label{ljmu}\\
\and School of Mathematical Sciences, Queen Mary, University of London, Mile End Road, London E1 4NS, UK \label{queenmary}\\
\and Vintage Lane Observatory, Blenheim, New Zealand \label{vintagelane}
\and Auckland Observatory, Auckland, New Zealand \label{aucklandobs}\\
\and Dept. of Physics and Astronomy, Texas A\&M University College Station, TX 77843-4242, USA  \label{texasam}\\
\and Possum Observatory, Patutahi, Gisbourne, New Zealand \label{possum}\\ 
\and Farm Cove Observatory, Centre for Backyard Astrophysics, Pakuranga, Auckland, New Zealand \label{farmcove}\\
\and Institute for Radiophysics and Space Research, AUT University, Auckland, New Zealand \label{aucklandaut}\\
\and Dept. of Astronomy and Space Science, Chungnam University, Korea\ \label{chungnam}\ 
\and Departamento de Astronomi{\'a} y Astrof{\'i}sica, Universidad de Valencia, E-46100 Burjassot, Valencia, Spain  \label{valencia}\\ 
\and UPMC-CNRS, UMR7095, Institut d'Astrophysique de Paris, 98bis boulevard Arago, F-75014, Paris, France \label{iap}\\ 
\and University of Canterbury, Dept. of Physics and Astronomy, Private Bag~4800, Christchurch 8020, New Zealand  \label{canterbury}\\
\and McDonald Observatory, 16120 St Hwy Spur 78 \#2, Fort Davis, Tx 79734, USA  \label{mcdonald}\\ 
\and University of the Free State, Faculty of Natural and Agricultural Sciences, Dept. of Physics, P.O.~Box 339, Bloemfontein 9300, South Africa  \label{freestate}\\ 
\and School of Math and Physics, University of Tasmania, Private Bag 37, GPO Hobart, Tasmania 7001, Australia  \label{utas}\\ 
\and Institute of Geophysics and Planetary Physics (IGPP), L-413, Lawrence Livermore National Laboratory, P.O. Box 808, Livermore, CA 94551, USA \label{llnl}\\ 
\and UniversitŽ de Toulouse; UPS-OMP; IRAP; Toulouse, France \label{latt}\\ 
\and CNRS; IRAP; 14, avenue Edouard Belin, F-31400 Toulouse, France \label{cnrstoulouse}\\
\and Physics Department, Faculty of Arts and Sciences, University of Rijeka, Omladinska 14, 51000 Rijeka, Croatia \label{rijeka}\\ 
\and Technical University of Vienna, Department of Computing, Wiedner Hauptstrasse 10, Vienna, Austria \label{tuv}\\ 
\and NASA Exoplanet Science Institute, Caltech, MS 100-22, 770 S.~Wilson Ave., Pasadena, CA 91125, USA \label{nexsci}\\ 
\and Perth Observatory, Walnut Road, Bickley, Perth 6076, Australia \label{perthobs}\\ 
\and South African Astronomical Observatory, P.O.~Box 9, Observatory 7935, South Africa \label{saao}\\ 
\and Solar-Terrestrial Environment Laboratory, Nagoya University, Nagoya, 464-8601, Japan \label{nagoya}\\
\and Dept. of Physics, University of Notre Dame, Notre Dame, IN~46556, USA \label{notredame}\\
\and Institute of Information and Mathematical Sciences, Massey University, Private Bag 102-904, North Shore Mail Centre, Auckland, New Zealand \label{massey}\\
\and Dept. of Physics, University of Auckland, Private Bag~92019, Auckland, New Zealand \label{auckland}\\
\and Okayama Astrophysical Observatory, National Astronomical Observatory of Japan, Asakuchi, Okayama 719-0232, Japan \label{okayama}\\
\and Mt.~John Observatory, P.O.~Box~56, Lake Tekapo 8770, New Zealand \label{mtjohn}\\
\and Dept. of Physics, Konan University, Nishiokamoto 8-9-1, Kobe 658-8501, Japan \label{konan}\\
\and Nagano National College of Technology, Nagano, 381-8550, Japan \label{nagano}\\
\and Tokyo Metropolitan College of Industrial Technology, Tokyo, 116-8523, Japan \label{tokyomet}\\
\and Dept. of Earth and Space Science, Graduate School of Science, Osaka University, 1-1 Machikaneyama-cho, Toyonaka, Osaka 560-0043, Japan \label{osaka}\\
\and Universidad de Concepci{\'o}n, Departamento de Astronomia, Casilla 160--C, Concepci{\'o}n, Chile \label{concepcion}
}

   \date{Received ... ; accepted ...}

 
  \abstract
   {}
   {We present the analysis of the gravitational microlensing event OGLE-2011-BLG-0251. This anomalous event was observed by several survey and follow-up collaborations conducting microlensing observations towards the Galactic Bulge.}
   {Based on detailed modelling of the observed light curve, we find that the lens is composed of two masses with a mass ratio $q=1.9 \times 10^{-3}$. Thanks to our detection of higher-order effects on the light curve due to the Earth's orbital motion and the finite size of source, we are able to measure the mass and distance to the lens unambiguously.}
   {We find that the lens is made up of a planet of mass $0.53 \pm 0.21\,\Mjup$ orbiting an M dwarf host star with a mass of $0.26 \pm 0.11 \Msun$. The planetary system is located at a distance of $2.57 \pm 0.61$ kpc towards the Galactic Centre. The projected separation of the planet from its host star is $d=1.408 \pm 0.019$, in units of the Einstein radius, which corresponds to $2.72 \pm 0.75$ AU in physical units. We also identified a competitive model with similar planet and host star masses, but with a smaller orbital radius of $1.50 \pm 0.50$ AU. The planet is therefore located beyond the snow line of its host star, which we estimate to be around $\sim 1-1.5$ AU.}
   {}

\titlerunning{A Cool Giant Planet in Microlensing Event OGLE-2011-BLG-0251}
   \maketitle
%

\section{Introduction}\label{sec:intro}
Gravitational microlensing is one of the methods that allows us to probe the populations of extrasolar planets in the Milky Way, and has now led to the discoveries of 16 planets\footnote{\tt http://exoplanet.eu}, several of which could not have been detected with other techniques (e.g. \citealt{beaulieu06}, \citealt{gaudi08}, \citealt{muraki11}). In particular, microlensing events can reveal cool, low-mass planets that are difficult to detect with other methods. Although this method presents several observational and technical challenges, it has recently led to several significant scientific results. \cite{sumi11} analysed short time-scale microlensing events and concluded that these events were produced by a population of Jupiter-mass free-floating planets, and were able to estimate the number of such objects in the Milky Way. \cite{cassan12} used 6 years of observational data from the PLANET collaboration to build on the work of \cite{gould10} and \cite{sumi11}, and derived a cool planet mass function, suggesting that, on average, the number of planets per star is expected to be more than 1.

Modelling gravitational microlensing events has been and remains a significant challenge, due to a complex parameter space and computationally demanding calculations. Recent developments in modelling methods \citep[e.g.][]{cassan08, kains09, kains12a, bennett10, ryu10, bozza12}, however, have allowed microlensing observing campaigns to optimise their strategies and scientific output, thanks to real-time modelling providing prompt feedback to observers as to the possible nature of ongoing events.

In this paper we present an analysis of microlensing event OGLE-2011-BLG-0251, an anomalous event discovered during the 2011 season by the OGLE collaboration and observed intensively by follow-up teams. In \Sec{sec:microlensing}, we briefly summarise the basics of relevant microlensing formalism, while we discuss our data and reduction in \Sec{sec:data}. Our modelling approach and results are outlined in \Sec{sec:modelling}; we translate this into physical parameters of the lens system in \Sec{sec:lensprop} and discuss the properties of the planetary system we infer.

\section{Microlensing formalism}\label{sec:microlensing}

Microlensing can be observed when a source becomes sufficiently aligned with a lens along the line of sight that the deflection of the source light by the lens is significant. A characteristic separation at which this occurs is the Einstein ring radius. When a single point source approaches a single point lens of mass $M$ with a projected source-lens separation $u$, the source brightness is magnified following a symmetric ``point source-point lens" (PSPL) pattern which can be parameterised with an impact parameter $\uz$ and a timescale $\te$, 
both expressed in units of the angular Einstein radius \citep{einstein36},

\begin{equation}
\label{eq:thetae}
	\thetae = \sqrt{ \frac{ 4\,G\,M }{ c^2 }
	\left( \frac{ D_\mathrm{S} - D_\mathrm{L} }
	{D_\mathrm{S}\, D_\mathrm{L} } \right)}
\ ,
\end{equation}

\noindent
where $G$ is the gravitational constant, $c$ is the speed of light, and $\ds$ and $\dl$ are the distances to the source 
and the lens, respectively, from the observer. The timescale is then $\te = \thetae/\mu$, where $\mu$ is the lens-source relative proper motion. Therefore the observable $\te$ is a degenerate function of $M, \dl$ and the source's transverse velocity $v_{\perp}$, assuming that $\ds$ is known. However, measuring certain second-order effects in microlensing light curves such as the parallax due to the Earth's orbit allows us to break this degeneracy and therefore measure the properties of the lensing system directly.

When the lens is made up of two components, the magnification pattern can follow many different morphologies, because of singularities in the lens equation. These lead to source positions, along closed \textit{caustic} curves, where the lensing magnification is formally infinite for point sources, although the finite size of sources means that, in practice, the magnification gradient is large rather than infinite. A point-source binary-lens (PSBL) light curve is often described by 6 parameters: the time at which the source passes closest to the center of mass of the binary lens, $\tz$, the Einstein radius crossing time, $\te$, the minimum impact parameter $\uz$, which are also used to describe PSPL light curves, as well as the source's trajectory angle $\alpha$ with respect to the lens components, the separation between the two mass components, $d$, and their mass ratio $q$. Finite source size effects can be parameterised in a number of ways, usually by defining the angular size of the source $\rho_*$ in units of $\thetae$:

\begin{equation}
\label{eq:thetas}
	\rho_* = \frac{\theta_*}{\thetae} \, ,
\end{equation}

\noindent
where $\theta_*$ is the angular size of the source in standard units.

\section{Observational data}\label{sec:data}

The microlensing event OGLE-2011-BLG-0251 was discovered by the OGLE (Optical Gravitational Lens Experiment) collaboration's Early Warning System \citep{udalski03} as part of the release of the first 431 microlensing alerts following the OGLE-IV upgrade. The source of the event has equatorial coordinates $\alpha=17^h38^m14.18^s$ and $\delta=-27^\circ08'10.1''$ (J2000.0), or Galactic coordinates of ($l, b$)=($0.670^\circ, 2.334^\circ$). Anomalous behaviour was first detected and alerted on August 9, 2011 (HJD$\sim$2455782.5) thanks to real-time modelling efforts by various follow-up teams that were observing the event, but by that time a significant part of the anomaly had already passed, with sub-optimal coverage due to unfavourable weather conditions. The anomaly appears as a two-day feature spanning HJD = 2455779.5 to 2455781.5, just before the time of closest approach $\tz$. Despite difficult weather and moonlight conditions, the anomaly was securely covered by data from five follow-up telescopes in Brazil ($\mu$FUN Pico dos Dias), Chile (MiNDSTEp Danish 1.54m) New Zealand ($\mu$FUN Vintage Lane, and MOA Mt. John B\&C), and the Canary Islands (RoboNet Liverpool Telescope). 

The descending part of the light curve also suffered from the bright Moon, with the source $\sim 5$ degrees from the Moon at $\sim 85\%$ of full illumination, leading to high background counts in images and more scatter in the reduced data. We opted not to include data from Mt. Canopus 1m telescope in the modelling because of technical issues at the telescope affecting the reliability of the images, and also excluded the $I$-band data from CTIO because they also suffer from large scatter, probably due to the proximity of the bright full Moon to the source.   

\begin{figure*}
  \centering
 \includegraphics[width=12cm, angle=0]{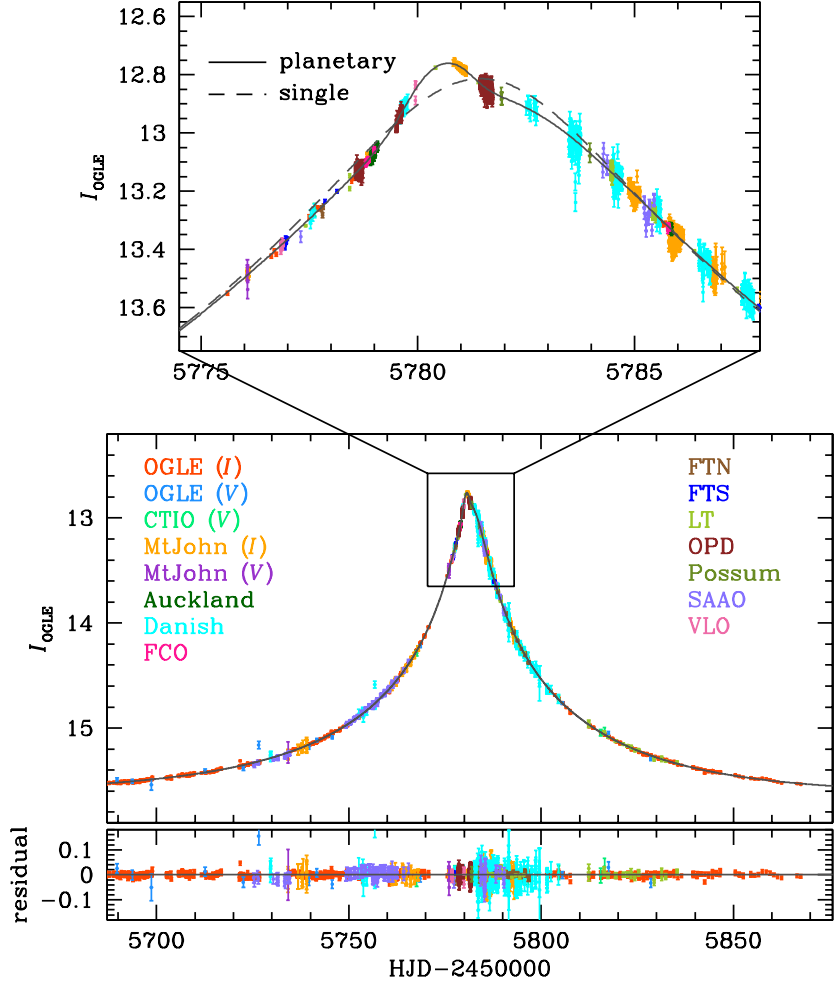}

  \caption{Light curve of OGLE-2011-BLG-0251. Data points are plotted with 1-$\sigma$ error bars, and the upper panel shows a zoom around the perturbation region near the peak \label{fig:lc}}

\end{figure*}

The data set amounts to 3738 images from 13 sites, from the OGLE survey team, the MiNDSTEp consortium, the RoboNet team, as well as the $\mu$FUN, PLANET and MOA collaborations in the $I, V$ and $R$ bands, as well as some unfiltered data; data sets are summarised in \Tab{tab:data} and the light curve is shown in \Fig{fig:lc}. We reduced all data using the difference imaging pipeline {\tt DanDIA} \citep{bramich08, bramich13}, except for the OGLE data, which was reduced by the OGLE team with their optimised offline pipeline.

\begin{table*}
\begin{center}
  \begin{tabular}{llllllll}
    \hline
 Team and telescope		& filter 		&Aperture		& Location				&$N$ 	&$a$	&$b$\\
    \hline
OGLE 					&$I$ 			&1.3m		& Las Campanas, Chile		& 1527 	&0.369	&0.020\\
OGLE 					&$V$  		&1.3m		& Las Campanas, Chile		& 27 		&0.937	&0.010\\
MiNDSTEp Danish 			&$I$ 			&1.54m		& La Silla, Chile			& 454 	&1.085	&0.020\\
LCOGT Liverpool Telescope 	&$I$			&2m			& La Palma, Canary Islands	& 191 	&2.434	&0.001\\
LCOGT Faulkes North 		&$I$	 		&2m			& Haleakala, Hawai'i		& 41 		&1.806	&0.005\\
LCOGT Faulkes South 		&$I$			&2m  & Siding Spring Observatory, Australia	& 31 		&1.119	&0.005\\
$\mu$FUN CTIO 			&$V$		&1.3m		& Cerro Tololo, Chile	    	& 6		&1.000	&0.020\\
$\mu$FUN Auckland 		&$R$		&0.4m		& Auckland, New Zealand	& 60 		&1.027	&0.010\\
$\mu$FUN Farm Cove 		&$-$			&0.36m		& Auckland, New Zealand	& 47 		&0.841	&0.005\\
$\mu$FUN Possum 			&$R$		&0.36m		& Gisborne, New Zealand	    	& 5 		&1.000	&0.020\\
$\mu$FUN Vintage Lane 		&$-$			&0.4m		& Blenheim, New Zealand	& 17 		&2.055	&0.001\\
$\mu$FUN Pico dos Dias		&$I$			&0.6m		& Minas Gerais, Brazil	    	& 572 	&3.095	&0.001\\
MOA Mt John B\&C 			&$I$			&0.6m		& South Island, New Zealand	& 621 	&5.175	&0.001\\
MOA Mt John B\&C 			&$V$		&0.6m		& South Island, New Zealand	& 5 		&1.000	&0.020\\
PLANET SAAO 			&$I$			&1m			& SAAO, South Africa    		& 134 	&1.931	&0.010\\
\hline
Total	&	&	&	&3738 \\

\hline
  \end{tabular}
  \caption{Data sets for OGLE-2011-BLG-0251, with the number of data points for each telescope/ filter combination. The rescaling coefficients $a$ and $b$ are also given, with error bars rescaled as $\sigma' = a\sqrt{\sigma^2 + b^2}$, where $\sigma'$ is the rescaled error bar and $\sigma$ is the original error bar. \label{tab:data}}
  \end{center}
\end{table*}

For each data set, we applied an error bar rescaling factors $a$ and $b$ to normalise error bars with respect to our best-fit model (see \Sec{sec:modelling}), using the simple scaling relation $\sigma_i' = a\sqrt{\sigma_i^2 + b^2}$ where $\sigma_i'$ is the rescaled error bar of the $i^{\rm th}$ data point and $\sigma_i$ is the original error bar. The error bar rescaling factors for each data set is given in \Tab{tab:data}. We did not exclude outliers from our data sets, unless we had reasons to believe that an outlier had its origin in a bad observation, or in issues with the data reduction pipeline.

\section{Modelling}\label{sec:modelling}

We modelled the light curve of the event using a Markov Chain Monte Carlo (MCMC) algorithm with adaptive step size. We first used the ``standard" PSBL parameterisation in our modelling, whereby a binary-lens light curve can be described by 6 parameters: those given in \Sec{sec:microlensing}, ignoring the second-order $\rho_*$ parameter described in that section. For all models and configurations we searched the parameter space for solutions with both a positive and a negative impact parameter $\uz$.

We started without including second-order effects of the source having a finite size or parallax due to the orbital motion of Earth around the Sun, and then added these separately in subsequent modelling runs by fitting the source size parameter $\rho_*$, as defined in \Sec{sec:microlensing}, and the parallax parameters described below. Both effects led to a large decrease in the $\chi^2$ statistic of the model ($> 1000$), which could not be explained only by the extra number of parameters. 

For the finite-source effect, we additionally considered the limb-darkening variation of the source star surface brightness by modelling the surface-brightness profile as

\begin{equation}\label{eq:lld}
I_{\psi, \lambda} = I_{0, \lambda} [1 - c_l\,(1-\cos\psi)] \, ,
\end{equation}

\noindent
where $I_{0, \psi}$ is the brightness at the centre of the source, and $\psi$ is the angle between a normal to the surface and the line of sight. We adopt the limb-darkening coefficients based on the source type determined from the dereddened colour and brightness (see \Sec{sec:sourceprop}). The values of the adopted coefficients are $c_V=0.073, c_I=0.624, c_R=0.542$, based on the catalogue of \cite{claret00}.

Finally, in a third round of modelling, we included both the effects of parallax and finite source size (``ESBL + parallax"). Including these effects together led to a significant improvement of the fit, with $\Delta \chi^2>500$ compared to the fits in which those effects were added separately. Computing the $f$-statistic (see e.g. \citealt{lupton93}) for this difference tells us that the probability of this difference occurring solely due to the number of degrees of freedom decreasing by 1 or 2 is highly unlikely. Our best-fit ESBL + parallax model is shown in \Fig{fig:lc}. 

To model the effect of parallax, we used the geocentric formalism \citep{dominik98a, an02, gould04}, which has the advantage of allowing us to obtain a good estimate of $\tz, \te$ and $\uz$ from a fit that does not include parallax. This formalism adds a further 2 parallax parameters, $\piee$ and $\pien$, the components of the lens parallax vector $\boldsymbol{\pie}$ projected on the sky along the east and north equatorial coordinates, respectively.  The amplitude of $\boldsymbol\pie$ is then

\begin{equation}\label{eq:pie}
\pi_E = \sqrt{\piee^2 + \pien^2}\, .
\end{equation}

Measuring $\pi_E$ in addition to the source size allows us to break the degeneracy between the mass, distance and transverse velocity of the lens system that is seen in \Eq{eq:thetae}. This is because $\pi_E$ also relates to the lens and source parallaxes $\pi_L$ and $\pi_S$ as

\begin{equation}\label{eq:pie2}
\pi_E = \frac{\pi_L - \pi_S}{\thetae} = \frac{\dl^{-1} - \ds^{-1}}{\thetae} .
\end{equation}

\noindent
Using this in \Eq{eq:thetae} allows us to solve for the mass of the lens.

As an additional second-order effect, we also consider the orbital motion of the binary lens. Under the approximation that the change rates of the binary separation and the rotation of the binary axis are uniform during the event, the orbital effect is taken into consideration with 2 additional parameters of $\dot{d}$ and $\dot{\alpha}$, which represent the rate of change of the binary separation and the source trajectory angle with respect to the binary axis, respectively. It is found that the improvement of fits by the orbital effect is negligible and thus our best-fit model is based on a static binary lens.

Below we outline our modelling efforts that resulted in fits that were not competitive with our best-fit ESBL + parallax models, and which we therefore excluded in our light curve interpretation.

\subsection{Excluded models}

\subsubsection{Xallarap}

We attempted to model the effects of so-called xallarap, orbital motion of the source if it has companion \citep{griesthu92}. Modelling this requires five additional parameters: the components of the xallarap vector, $\xi_{\rm{E}, N}$ and $\xi_{\rm{E}, E}$, the orbital period $P$, inclination $i$ and the phase angle $\psi$ of the source orbital motion. By definition, the magnitude of the xallarap vector is the semi-major axis of the source's orbital motion with respect to the centre of mass, $a_{\rm S}$, normalised by the projected Einstein radius onto the source plane, $\hat{r}_{\rm E}=\ds \thetae$, i.e. 

\begin{equation}\label{eq:xall_xie}
\xi_{\rm E}=a_{\rm S}/\hat{r}_{\rm E}\, .
\end{equation}

\noindent
The value of $a_{\rm S}$ is then related to the semi-major axis of the binary by

\begin{equation}\label{eq:xall_a}
a_{\rm S}=\frac{a\,M_2}{M_1 + M_2}\, ,
\end{equation}

\noindent
where $M_1$ and $M_2$ are the masses of the source components. 

\begin{figure}
  \centering
 \includegraphics[width=8cm, angle=0]{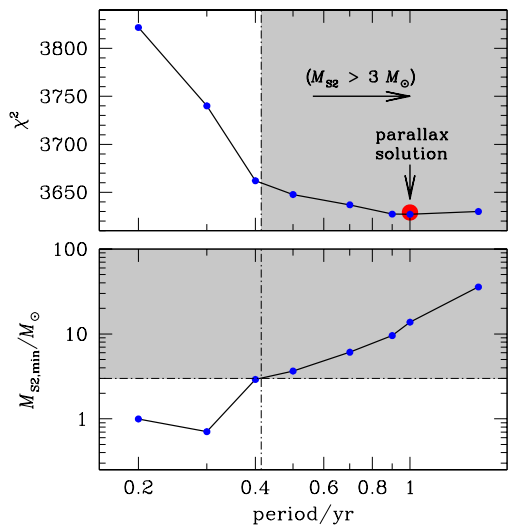}
  \caption{Constraints from the xallarap fit as a function of the orbital period $P$ of the source star. The top panel shows $\chi^2$ of the xallarap fit as a function of $P$, with a red circle marking the location of the best parallax model. The bottom panel shows the minimum mass of the source companion as a function of $P$. The shaded area in both panels indicates where models are excluded based on conservative blending constraints on the source companion's mass. \label{fig:xall_constraint}}

\end{figure}

In \Fig{fig:xall_constraint}, we show $\chi^2$ of the fit plotted as a function of the orbital period of the source star. We compare this to the $\chi^2$ statistic of the best parallax fit. We find that xallarap models provide fits competitive with the parallax planetary models for orbital periods $P>1$ year. However, the solutions in this range cannot meet the constraint provided by the source brightness. Combining Equations (\ref{eq:xall_xie}) and (\ref{eq:xall_a}) with Kepler's third law, $P^2=a^3/(M_1+M_2)$ yields \citep{dong09}

\begin{equation}\label{eq:xall_kepler}
P^2=\frac{(M_1 + M_2)^2}{M_2^3}\left(\frac{\xi_{\rm E}\hat{r}_{\rm E}}{\rm AU}\right)^3\, .
\end{equation}

\noindent
Rearranging this equation for $M_2$, and using the fact that $M2/(M_1 + M_2) < 1$, we can derive an upper limit for the mass of $M_2$, 

\begin{equation}\label{eq:xall_m2min}
M_{2, \rm min} =\frac{(\xi_{\rm E}\hat{r}_{\rm E})^3}{P^2}\, .
\end{equation}

In the lower panel of \Fig{fig:xall_constraint}, we show the minimum mass of the source companion as a function of orbital period. The blending constraint means that the source companion cannot be arbitrarily massive, and we use a conservative upper limit for its mass of 3 $\msun$. With this constraint, we find that xallarap models are not competitive with parallax planetary models, and we therefore exclude the xallarap interpretation of the light curve.

\subsubsection{Binary source}

We also attempted to model this event as a binary source - point lens (BSPL) event. For this we introduced three additional parameters: the impact parameter of the secondary source component, $u_{0, 2}$, and its time of closest approach, $t_{0, 2}$, as well the flux ratio between the source components. We note that parallax is also considered in our binary source modelling, for fair comparison to other models. We find that the best binary-source model provides a poorer fit, with $\chi^2=3809$, which gives $\Delta \chi^2 \sim 180$ compared to our best planetary model (including parallax, see model D in the following section). Residuals for this model, as well as all other models discussed in this section are shown in \Fig{fig:residuals}.

\subsection{Best-fit models}

\begin{figure*}
  \centering
 \includegraphics[width=12cm, angle=0]{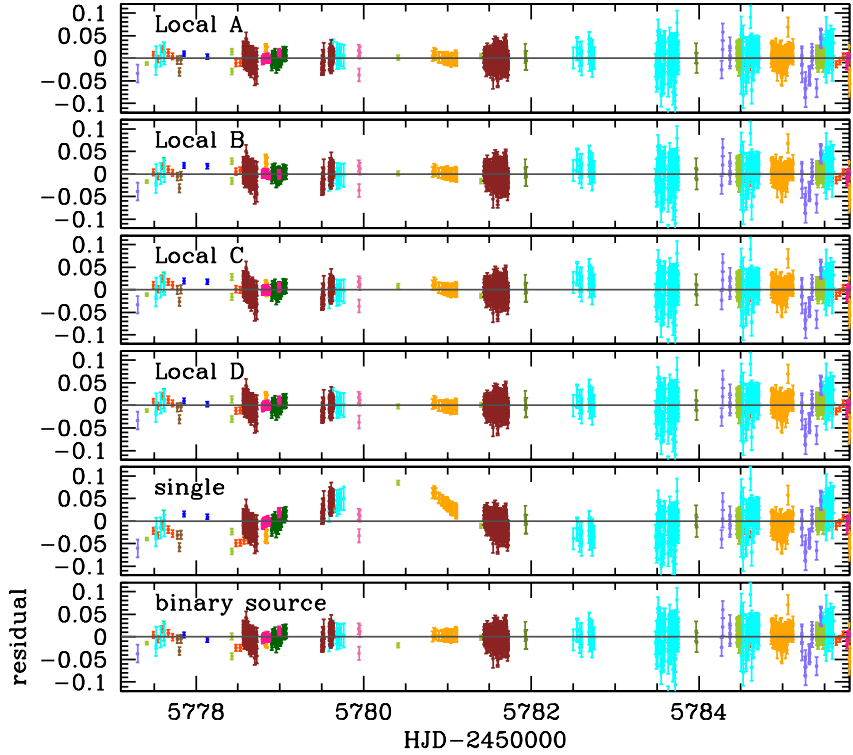}

  \caption{Residual of data, with 1-$\sigma$ error bars, for the various models considered. \label{fig:residuals}}

\end{figure*}

\begin{figure*}
  \centering
  \includegraphics[width=12cm, angle=0]{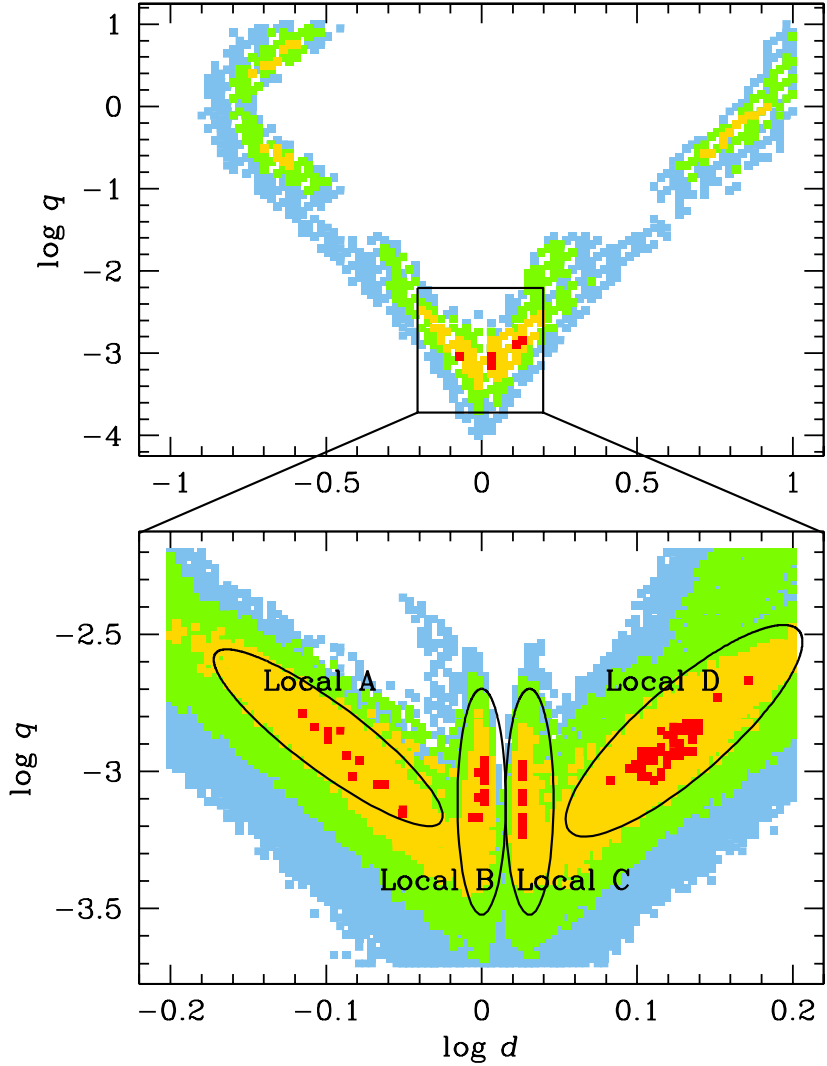}

  \caption{$\chi^2$ map in the $d, q$ plane, showing the location of the four local minima identified by our modelling runs. Out of these, local minima $A$ and $D$ are competitive, with local minima $B$ and $C$ having $\Delta \chi^2 \sim 50$ and 70 respectively, for the same number of parameters. Minima $A$ and $D$ correspond to the close and wide ESBL + parallax models discussed in the text. Different colours correspond to $\Delta\chi^2<$ 25 (red), 100 (yellow), 225 (green), and 400 (blue); we note that the $\chi^2$ map is based on the original data, before error-bar normalisation, and therefore the $\Delta\chi^2$ levels are slightly different from those given in \Tab{tab:par_allmodels}. The top panel shows the breadth of our parameter space exploration, encompassing planetary and non-planteray mass-ratio regimes, while the bottom panel shows a zoom on the region where our local minima are located.\label{fig:hanmap}}

\end{figure*}

We searched the parameter space using an MCMC algorithm as well as a grid of $(d, q, \alpha)$ to locate good starting points for the algorithm (see e.g. \citealt{kains09}), over the range $-4 < \log q < 0$ and $-1.0 < \log d < 2$. This encompasses both planetary and binary companions that might cause the central perturbation. In \Fig{fig:hanmap} we present the $\chi^2$ distribution in the $d, q$ plane. We find four local solutions, all of which have a mass ratio corresponding to a planetary companion. We designate them as A, B, C and D; the degeneracy among these local solutions is rather severe, as can be seen from the residuals shown in \Fig{fig:residuals}.

\begin{figure*}
  \centering
 \includegraphics[width=12cm, angle=0]{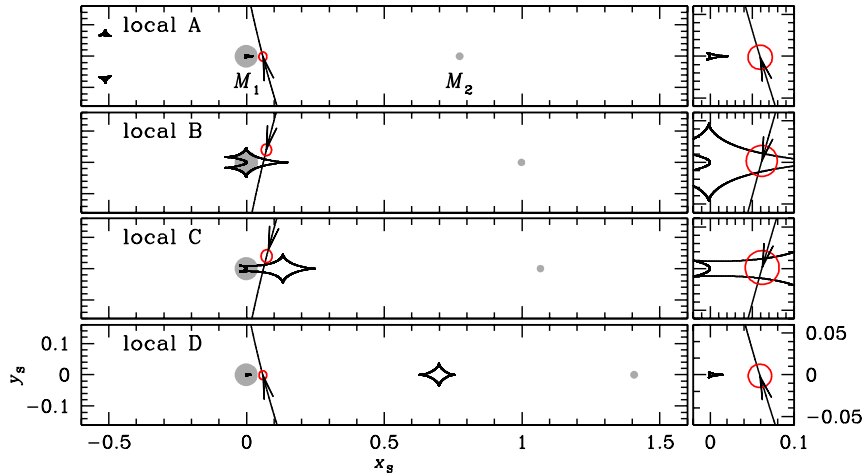}
\caption{Source trajectory geometry with respect to the caustics for all four local minima identified in \Fig{fig:hanmap}; the source size is marked as a red circle. \label{fig:geo}}
\end{figure*}

\begin{figure*}
  \centering
  \includegraphics[width=13cm, angle=0]{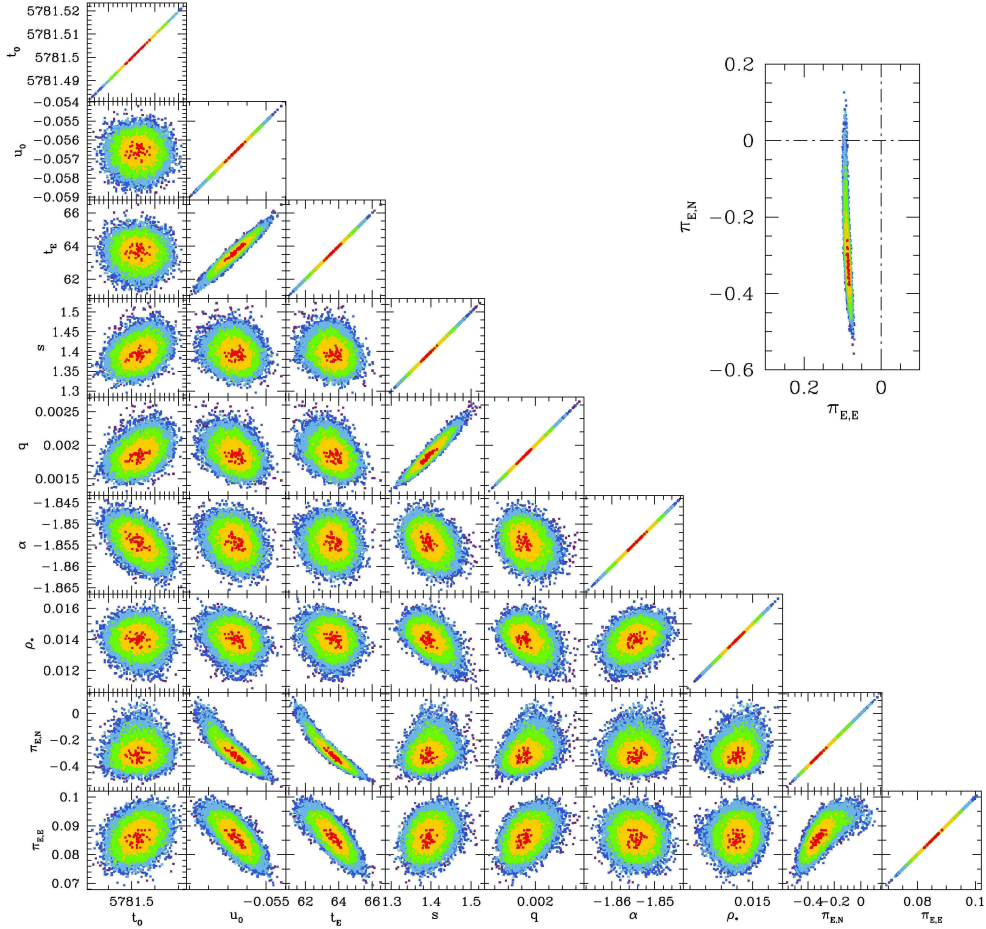}

  \caption{Parameter-parameter correlations for our 9 fitted parameters. Colours indicate the limits of the 1, 2, 3, 4 and 5-$\sigma$ confidence limits for each pairwise distribution. A closer view of the correlation between parallax parameters is shown on the top right inset. \label{fig:correl}}

\end{figure*}

For the identified local minima, we then further refine the lensing parameters by conducting additional modelling, considering higher-order effects of the finite source size and the EarthÕs orbital motion. It is found that the higher-order effects are clearly detected with $\Delta\chi^2 > 500$. Best-fit parameter for each of the local minima are given in \Tab{tab:par_allmodels}, while \Fig{fig:geo} shows the geometry of the source trajectories with respect to the caustics for all four minima. We note that the pairs of solutions A and D, and B and C, are degenerate under the well-known $d \leftrightarrow d^{-1}$ degeneracy \citep{griest98, dominik99a}; this is caused by the symmetry of the lens mapping between binaries with $d$ and $d^{-1}$. Comparing the pairs of solutions, we find that the A-D pair is favoured, with $\Delta\chi^2 > 40$ compared to the B-C pair. On the other hand, the degeneracy between the A and D solutions is very severe, with only $\Delta\chi^2 \sim 7$. In \Fig{fig:correl}, we also show parameter-parameter correlations plots for model D, showing also the uncertainties in the measured lensing parameters.

\begin{table*}
\begin{center}
  \begin{tabular}{cccccc}
    \hline
    Parameter 	& Local A			& Local B		& Local C 		& Local D 	\\
    \hline
$\chi^2$		&3636			&3698		&3675	&3629 \\
$d.o.f.$		&3699			&3699		&3699	&3699 \\
\hline    
$\tz$ [MHJD] &5781.509 $\pm$ 0.004 	&5781.472 $\pm$ 0.004	&5781.487 $\pm$ 0.004	 	&5781.503 $\pm$ 0.004\\
$\te$ [days] &  63.74 $\pm$ 0.41 		& 64.05 $\pm$ 0.46 		&64.24 $\pm$ 0.47			&  63.88 $\pm$ 0.46\\
$\alpha$ [rad] &   -1.855 $\pm$ 0.002 	& -1.845 $\pm$ 0.002 	&-1.849 $\pm$ 0.004		&  -1.855 $\pm$ 0.002\\
$\uz/10^{-2}$ 	&   -5.66 $\pm$ 0.04 			& -5.63 $\pm$ 0.04	 	&-5.64 $\pm$ 0.05			&   -5.63 $\pm$ 0.04\\
$\rho_*/10^{-2}$ 	&  1.44 $\pm$ 0.05			&   1.77 $\pm$ 0.04	 	&1.87 $\pm$ 0.08			& 1.39 $\pm$ 0.05\\
$d$ 		&   0.775 $\pm$ 0.010 		&   0.997 $\pm$ 0.009	&$1.066 \pm 0.001$	 		&  1.408 $\pm$ 0.019\\
$q/10^{-3}$ &(1.68 $\pm$ 0.11) 		&(0.93 $\pm$ 0.03) 		&(1.11 $\pm$ 0.06) 			&(1.92 $\pm$ 0.12)\\
$\pien$ 	&  -0.33 $\pm$ 0.04			&  -0.37 $\pm$ 0.04 		&$-0.40 \pm 0.05$			&  -0.34 $\pm$ 0.05\\	
$\piee$ 	&  0.09 $\pm$ 0.01			& 0.08 $\pm$ 0.01		&$0.08 \pm 0.01$			&  0.09 $\pm$ 0.01\\
$\pie$ 	&  0.34 $\pm$ 0.04 			& 0.38 $\pm$ 0.04 		&$0.41 \pm 0.05$			&  0.35 $\pm$ 0.05\\
$^a g=\fb/\fs$    &   0.387 $\pm$ 0.035 	&0.394 $\pm$ 0.001 	&0.394 $\pm$ 0.042			&   0.376 $\pm$ 0.017\\
$^a I_S$		&   15.99 $\pm$  0.03 	&15.98 $\pm$  0.01 		&15.98 $\pm$ 0.02			&   15.97 $\pm$  0.01\\
$^a I_B^a$ 	&   16.97 $\pm$  0.07 	&16.99 $\pm$  0.01 		&16.99 $\pm$ 0.06			&   17.04 $\pm$  0.03\\

\hline
  \end{tabular}
  \caption{Best-fit model parameters and 1-$\sigma$ error bars for the four identified best binary-lens models including the effects of the orbital motion of the Earth (parallax). MHJD$\equiv$HJD-2450000. $^a$for the OGLE data set \label{tab:par_allmodels}}
  \end{center}
\end{table*}

\section{Lens Properties}\label{sec:lensprop}

In this section we determine the properties of the lens system, using our best-fit model parameters, i.e. our wide-configuration ESBL + parallax model. We also calculated the lens properties for the competitive close-configuration model, with both sets of parameter values listed in \Tab{tab:lensprop}. 

\subsection{Source star and Einstein radius}\label{sec:sourceprop}

\begin{figure*}
  \centering
  \includegraphics[width=14cm, angle=0]{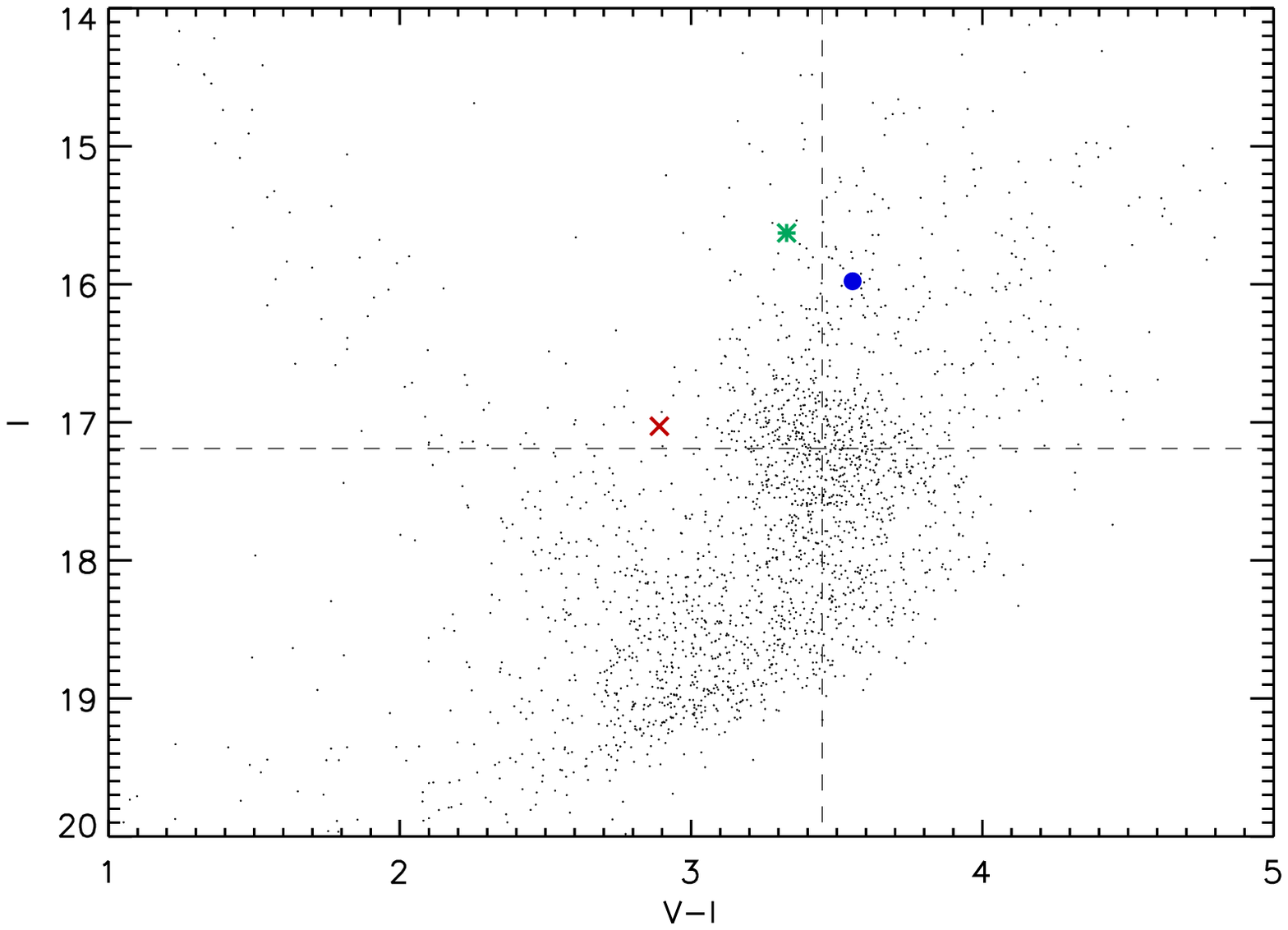}

  \caption{$V-I, I$ colour-magnitude diagram of the OGLE-2011-BLG-0251 field obtained using OGLE-IV photometry. The location of the total source + blend is indicated by a green asterisk, while the location of the deblended source is marked by a blue filled circle, and that of the blend by a red cross. The dashed lines cross at the location of the Red Clump. \label{fig:cmd}}

\end{figure*}

We determined the Einstein radius by first calculating the angular size of the source. This can be done by using the magnitude and colour of the source \citep[e.g.][]{yoo04a}, and empirical relations between these quantities and the angular source size. We start by using the location of the red giant clump (hereafter RC) on our colour-magnitude diagram (Fig. \ref{fig:cmd}) to estimate the reddening and extinction along the line of sight. We use an $I$-band absolute magnitude for the RC of $M_{I, {\rm RC}, 0}=-0.12\pm0.09$ \citep{nataf12}, as well as a colour $(V-I)_{{\rm RC}, 0} = 1.06\pm0.12$ \citep{bensby11}. We compare these values to those on our colour-magnitude diagram (CMD), which we generated using OGLE $I-$ and $V-$ band photometry. From \Fig{fig:cmd}, the location of the RC on our CMD is 

\begin{equation}
(I, V-I)_{\rm RC} = (17.19 \pm 0.05, \,3.45 \pm 0.05)
\end{equation}

so, using a distance modulus of $\mu=14.52 \pm 0.09$, i.e. a distance to the Galactic bulge of $8.0 \pm 0.3$ kpc \citep{yelda11}, we find $A_I=2.79 \pm 0.10$ and $E(V-I)=2.39 \pm 0.13$.

Using these values, the best-fit value for the magnitude of the source $I_S = 15.97 \pm 0.01$, a source colour $(V-I)_{S, 0}=1.15$, and the empirical relations of \cite{kervella08}, we find an angular source radius $\theta_* = 10.41 \pm 1.18 \, \uas$, or a source star radius of $R_* = 10.53 \pm 1.19 \, R_{\bigodot}$. This, together with the best-fit value of the source size parameter $\rho_*$, allows us to calculate the size of the Einstein radius, $\thetae = \theta_*/\rho_*$. Using the relevant parameter values, we find $\thetae=0.749 \pm 0.283$ mas. This in turn allows us to calculate the source-lens relative proper motion, $\mu_{\rm rel}=\thetae/\te = 4.28 \pm 1.62$ mas/yr.

\subsection{Masses of the Lens Components}\label{sec:lensmass}

Combining \Eq{eq:thetae} and \Eq{eq:pie2} allows us to derive an expression for the mass as a function of the parallax vector magnitude $\pi_E$ (defined by Eq. \ref{eq:pie}): 

\begin{equation}\label{eq:mpie}
M_L=\frac{\thetae c^2}{4G\pi_E}
\end{equation}

Using values found in the previous section, and our best-fit parallax parameter value $\pi_E=0.35\pm 0.05$ yields a total lens mass $M_L=0.26 \pm 0.10\,\msun$. Using the best-fit mass ratio parameter value of $q=(1.92 \pm 0.12) \times 10^{-3}$ yields component masses of $0.26 \pm 0.11\,\msun$ and $0.53 \pm 0.21\,\Mjup$, where $\Mjup$ is the mass of Jupiter.

\subsection{Distance to the Lens}

We can also rearrange \Eq{eq:thetae} to derive an expression for the distance to the lens $\dl$,

\begin{equation}\label{eq:dl}
\dl=\left[\frac{1}{\ds} + \frac{\thetae^2c^2}{4GM}\right]^{-1}\, .
\end{equation}

\noindent
Using our parameter values as well as the lens mass derived thanks to our parallax measurement, we find a distance to the lens of $\dl=2.57 \pm 0.61$ kpc. This distance allows us to carry out a sanity check of the lens mass we derived in the previous section. By assuming that the contribution from the blended light comes from the lens, we can derive an upper limit to the $I-$band lens magnitude $M_I$ using our best-fit blending parameter:

\begin{equation}\label{eq:absmag}
M_{I, {\rm L}}=m_{I, b} - 5\log_{10}\dl - 10 - A_{I, {\rm L}}\, ,
\end{equation}

\noindent
where $m_{I, b}$ is the apparent $I-$band magnitude of the blend, $A_{I, {\rm L}}$ is the extinction between the observer and the lens, and $\dl$ is in kpc. In practice, $A_{I, {\rm L}} \leq A_I$ since the lens is in front of the source, so we use the extreme scenario where $A_{I, {\rm L}}=A_I$ to derive an upper brightness limit (lower limit on the magnitude) for the lens. We find this to be $M_{I, \rm L} = 2.19 \pm 0.53$ mag, which corresponds to a maximum mass of the lens of $M_{L, \rm max} = 1.65 \pm 0.23 \Msun$, assuming a main sequence star mass-luminosity relation, and assuming that the secondary lens component (i.e. the planet) does not contribute to the blended light. This is much larger than the value we derived in \Sec{sec:lensmass} for the mass of the primary lens component, which suggests that some blending comes from stars near the source rather than from the lens, although it is difficult to quantify this without an estimate of $A_{I, \rm L}$.

Finally, we can also use the distance to the lens and the size of the Einstein ring radius to calculate the projected separation $r_{\perp}$ between the lens components in AU. Using our best-fit projected angular separation $d=1.408 \pm 0.019$, we find a projected (i.e. minimum) orbital radius $r_{\perp}=2.72 \pm 0.75$ AU. 

We can compare this to an estimate of the location of the ``snow line", which is the location at which water sublimated in the midplane of the protoplanetary disk, i.e. the distance at which the midplane had a temperature of $T_{\rm mid}=170$ K (although other studies have noted that this temperature varies; see e.g. \citealt{podolak04}). The core accretion model of planet formation predicts that giant planets form much more easily beyond the snow line, thanks to easier condensation of icy material and therefore easier formation of large solid cores in the early stages of the circumstellar disk's evolution. \cite{kennedy08} modelled the evolution of the snow line's location, taking into account heating of the disk via accretion, as well as the influence of pre-main sequence stellar evolution. Using a rough extrapolation of their results, we estimate that the snow line (at $t=1$Myr) for the planetary host star in OGLE-2011-BLG-0251 is located at around $\sim 1-1.5$ AU. We therefore conclude that OGLE-2011-BLG-0251Lb is a giant planet located beyond the snow line, with both of our competitive best-fit models yielding projected orbital radii larger than $1.5$ AU.

We list all the lens properties in \Tab{tab:lensprop}, both for the best-fit model parameters that we have used above, and for the close-configuration model, for comparison. Lens properties derived using the close-configuration model are very similar to those we found using the wide-configuration model, the only major difference being in the orbital radius. For the close model, we find an orbital radius of $1.50 \pm 0.50$ AU, which is close to the location of the snow line.

\begin{table}
\begin{center}
  \begin{tabular}{ccc}
    \hline
	&close 		&wide\\
    \hline
$\theta_*$ [${\mu}$as]    			&10.29 $\pm$1.17 	&10.41 $\pm$1.18 \\
$\thetae$ [mas] 				&0.71 $\pm$0.26 	&0.75 $\pm$0.28 \\
$\mu_{\rm rel}$ [mas yr$^{-1}$]		&4.09 $\pm$1.50 	&4.28 $\pm$1.62 \\
$M_1$ [$\Msun$]   				&0.26 $\pm$0.10 	&0.26 $\pm$0.11 \\
$M_2$ [$\Mjup$]   				&0.45 $\pm$0.18 	&0.53 $\pm$0.21 \\
$M_{L, \rm max}$       			&1.71 $\pm$0.23 	&1.65 $\pm$0.23 \\
$\dl$ [kpc]     					&2.71 $\pm$0.61 	&2.57 $\pm$0.61 \\
$r_{\perp}$ [AU]        				&1.50 $\pm$0.50 	&2.72 $\pm$0.75 \\

\hline
  \end{tabular}
  \caption{Lens properties derived as detailed in \Sec{sec:lensprop}, for both competitive parallax models. \label{tab:lensprop}}
  \end{center}
\end{table}

\section{Conclusions}

Our coverage and analysis of OGLE-2011-BLG-0251 has allowed us to locate and constrain a best-fit binary-lens model corresponding to an M star being orbited by a giant planet. This was possible through a broad exploration of the parameters both in real time, thanks to the recent developments in microlensing modelling algorithms, and after the source had returned to its baseline magnitude. Various second-order effects, as well as other possible, non-planetary, interpretations for the anomaly were considered during the modelling process. Based on the best-fit solution, we were able to constrain the masses and separation of the lens components, as well as various other characteristics, thanks to a strong detection of parallax effects due to the Earth's orbit around the Sun, in conjunction with the detection of finite source size effects. We found a planet of mass $0.53 \pm 0.21\, \Mjup$ orbiting a lens of $0.26 \pm 0.11 \Msun$ at a projected radius $r_{\perp} = 2.72 \pm 0.75$ AU; the whole system is located at a distance of $2.57 \pm 0.61$ kpc. Our competitive second-best model leads to similar properties, but a smaller projected orbital radius $r_{\perp}=1.50 \pm 0.50$. The two best-fit models are competitive and therefore we cannot make a strong claim about which orbital radius is favoured. However, comparing both values of the projected orbital radius to the approximate location of the snow line for a typical star of the mass of the primary lens component, we conclude that OGLE-2011-BLG-0251Lb is a giant planet located around or beyond the snow line. This is in line with predictions from the core accretion model of planet formation, from which we expect large planets to be more numerous beyond the snow line; this is also where microlensing detection sensitivity is at its highest, enabling us to probe a region of planetary parameter space that is difficult to reach for other methods.

\section*{Acknowledgements}
NK acknowledges an ESO Fellowship. The research leading to these results has received funding from the European Community's Seventh Framework Programme (/FP7/2007-2013/) under grant agreements No 229517 and 268421. 
The OGLE project has received funding from the European Research Council under the European Community's Seventh Framework Programme (FP7/2007-2013) / ERC grant agreement no. 246678 to AU. 
KA,DB,MD,KH,MH,SI,CL,RS,YT are supported by NPRP grant NPRP-09-476-1-78 from the Qatar National Research Fund (a member of Qatar Foundation).
Work by C. Han was supported by Creative Research Initiative Program (2009- 0081561) of National Research Foundation of Korea. 
This work is based in part on data collected by MiNDSTEp with the Danish 1.54m telescope at the ESO La Silla Observatory. The Danish 1.54m telescope is operated based on a grant from the Danish Natural Science Foundation (FNU). 
The MiNDSTEp monitoring campaign is powered by ARTEMiS (Automated Terrestrial Exoplanet Microlensing Search; \citealt{dominik08}). 
MH acknowledges support by the German Research Foundation (DFG). 
DR (boursier FRIA), OW (aspirant FRS - FNRS) and J. Surdej acknowledge support from the Communaut\'{e} fran\c{c}aise de Belgique -- Actions de recherche concert\'{e}es -- Acad\'{e}mie universitaire Wallonie-Europe. 
TCH gratefully acknowledges financial support from the Korea Research Council for Fundamental Science and Technology (KRCF) through the Young Research Scientist Fellowship Program. TCH and CUL acknowledges financial support from KASI (Korea Astronomy and Space Science Institute) grant number 2012-1-410-02.
Work by J.C. Yee is supported by a National Science Foundation Graduate Research Fellowship under Grant No. 2009068160. A. Gould and B.S. Gaudi acknowledge support from NSF AST-1103471. B.S. Gaudi, A. Gould, and R.W. Pogge acknowledge support from NASA grant NNX12AB99G.
The MOA experiment was supported by grants JSPS22403003 and JSPS23340064. TS was supported by the grant JSPS23340044. Y. Muraki acknowledges support from JSPS grants JSPS23540339 and JSPS19340058.

\bibliographystyle{aa}
\bibliography{../thesisbib}

\label{lastpage}

\end{document}